\documentclass[cameraready]{Interspeech}

\title{Emotion-Aware Quantization for Discrete Speech Representations: An Analysis of Emotion Preservation}

\author[affiliation={1}, equalcontribution]{Haoguang}{Zhou}
\author[affiliation={1}, equalcontribution]{Siyi}{Wang}
\author[affiliation={2}]{Jingyao}{Wu}
\author[affiliation={3}]{James}{Bailey}
\author[affiliation={1}]{Ting}{Dang}

\address{
    $^1$ The University of Melbourne, Australia, 
    $^2$ Massachusetts Institute of Technology, USA,\\
    $^3$ Monash University, Australia
}

\email{haoguang.zhou@student.unimelb.edu.au, siyi.wang.4@student.unimelb.edu.au}


\usepackage{comment}
\usepackage{multirow} 
\usepackage{soul, xcolor}
\usepackage{amsmath}
\usepackage{graphicx}
\usepackage{subcaption}
\usepackage[utf8]{inputenc}

\begin{document}

\maketitle

\begin{abstract}
Modern speech systems increasingly use discretized self-supervised speech representations for compression and integration with token-based models, yet their impact on emotional information remains unclear. 
We study how residual vector quantization (RVQ) reshapes emotional information in discrete speech representations from both representation- and task-level perspectives. 
Our analysis shows that aggressive compression disproportionately degrades emotion, with uneven loss across emotion classes and model architectures. 
To address this, we introduce emotion-aware quantization using emotion-specific and emotion-biased codebooks, improving the preservation of both hard and soft emotion perception. 
We further propose Emo-Q, a lightweight routed quantization method that selects emotion-specialized codebooks, improving emotion recognition performance at lower bitrates. 
These results highlight the importance of emotion-aware discretization for robust affective speech processing.
\end{abstract}

\section{Introduction}

Modern speech systems increasingly rely on discretized self-supervised representations for efficient storage and integration with token-based models \cite{borsos2023audiolm, mousavi2025discrete}. Contemporary self-supervised learning (SSL) front-ends \cite{hsu2021hubert, ma2024emotion2vec} encode speech into high-dimensional embeddings converted into discrete units via residual vector quantization (RVQ) \cite{defossez2022high}. However, discretization introduces an information bottleneck whose impact on affective information remains insufficiently understood. Unlike phonetic content, emotion is subtle and distributed across acoustic dimensions, making it vulnerable to representational distortion under compression \cite{ren2024emo, sun2026recovering}. While continuous SSL spaces are well-explored through layer-wise probing \cite{pepino2021emotion}, their organization within discretized spaces remains unclear, posing a challenge for developing efficient, affect-aware speech technologies.

\begin{figure}
  \centering
  \includegraphics[width=0.9\linewidth]{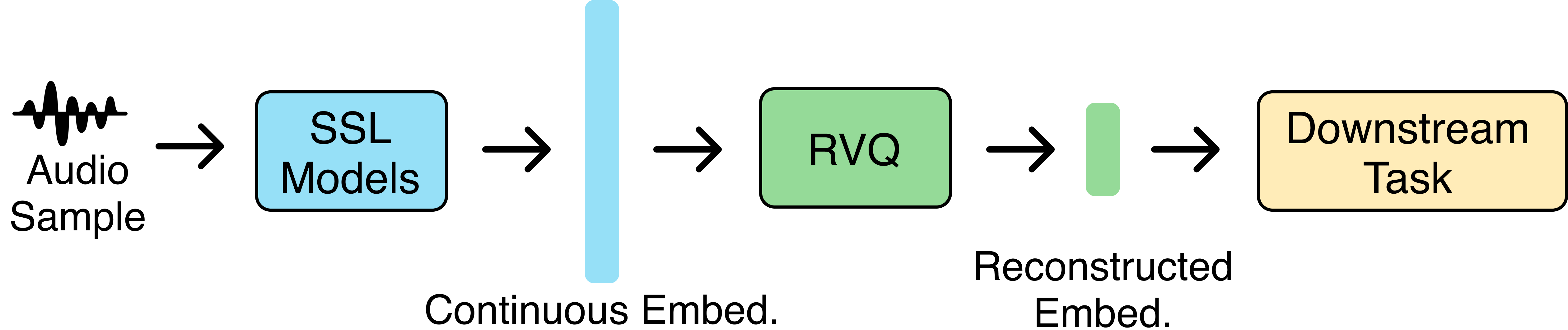}
  \caption{Quantization of discrete speech representation}
  \vspace{-20pt}
  \label{fig:1}
\end{figure}

Rather than viewing discretization as an inevitable information loss, we investigate whether the bottleneck can be structurally optimized. By framing the codebook as a task-sensitive tokenizer (Fig~\ref{fig:1}), we evaluate affective preservation across both \emph{representation} and \emph{task} levels via four research questions:

\textbf{RQ1 (Representation-level: Affective Information Decay)} How is emotional information structured and degraded under RVQ across different SSL architectures? We analyze the robustness of affective cues under increasing quantization depth. 

\textbf{RQ2 (Representation-level: Categorical Preservation)} Can emotion-specific codebook training improve affective preservation?
We study whether quantization can be designed to better capture emotional structure in representation space. 

\textbf{RQ3 (Representation-level: Distributional Fidelity)} Does emotion-specific quantization preserve fine-grained affective distributions?
Since real-world emotional expression is often ambiguous \cite{cabanac2002emotion}, we move beyond categorical emotion recognition to evaluate soft affective distribution preservation.

\textbf{RQ4 (Task-level: Downstream Utility)}: Can emotion-specific quantization improve downstream speech emotion recognition (SER) performance? Finally, we assess how representation-level gains translate into practical utility by improving downstream SER task performance. 
In summary, this work systematically studies affective information preservation under progressive quantization and proposes emotion-aware codebook specialization for affect-preserving compression. Results show that affect-aware quantization better preserves both categorical and soft affective structures, and can improve downstream speech emotion recognition performance under compressed representations. This study provides a general framework for affect-aware representation compression in speech intelligence systems.

\section{Related Work}

Vector quantization produces discrete units for token-based modeling \cite{zeghidour2021soundstream, defossez2022high}, but creates an information bottleneck that attenuates paralinguistic cues. 
Recent studies characterize this bottleneck by evaluating information accessibility in discretized HuBERT \cite{yeh2024estimating} and the downstream impact of representation inconsistency \cite{liu2025analyzing}. While these works confirm that discrete units retain certain cues \cite{mousavi2025discrete,nguyen2023expresso,ren2024emo}, they lack a systematic quantification of the efficiency–fidelity trade-off for emotion. Existing mitigation relies on auxiliary fusion \cite{sun2026recovering} or complex tokenizer redesign \cite{zhang2023speechtokenizer, sanders2025segmentation}, increasing modeling overhead. We thus investigate whether the bottleneck can be structurally optimized to preserve affective nuances without additional complexity.
\vspace{-5pt}

\section{Experimental Setup}
\label{sec:exprimental_setup}
\vspace{-2pt}
Code for reproducing the experiments is available:
\url{https://anonymous.4open.science/r/OOOOO-3C90/README.md}.
\vspace{-8pt}
\subsection{Data and Quantization}
\vspace{-3pt}
\label{sec:codebook-config}
\noindent\textbf{Datasets. }
All experiments use four English emotion speech corpora: ESD-EN~\cite{zhou2021seen}, IEMOCAP~\cite{busso2008iemocap}, RAVDESS~\cite{livingstone2018ryerson}, and CREMA-D~\cite{cao2014crema}, with four emotion categories (\textit{angry}, \textit{happy}, \textit{neutral}, \textit{sad}). For IEMOCAP, \textit{excited} is merged into \textit{happy} following standard practice. 
Each corpus is split 50/10/40\% for train/val/test.

For codebook evaluation in RQ1 and RQ2, we adopt a
leave-one-corpus-out protocol.
For RQ3, IEMOCAP is reserved entirely for evaluation;
as it provides per-annotator vote distributions provide ground-truth soft labels 
(e.g., 60\% angry, 30\% neutral, 10\% sad).
For RQ4, the test splits of the four training corpora are used for cross-validation of RVQ layer selection. We additionally evaluate on four OOD test sets:
MSP-Podcast~\cite{lotfian2017building} and three CAMEO
subsets~\cite{christop2025cameocollectionmultilingualemotional}
(EMNS~\cite{emns}, eNTERFACE~\cite{enterface}, JL-Corpus~\cite{jlcorpus}).

\noindent\textbf{SSL Front-ends. } For the emotion-targeted encoder, we use emotion2vec+ Base~\cite{ma2024emotion2vec}, a speech emotion representation model trained via self-supervised online distillation on unlabeled emotional speech, serving as a reference for emotion-focused quantization analysis. For general-purpose encoders, we use HuBERT-Large~\cite{hsu2021hubert} and WavLM-Large~\cite{chen2022wavlm}, two widely adopted SSL encoders trained on masked acoustic prediction without emotion-specific supervision.

\noindent\textbf{Residual Vector Quantization (RVQ). }

We aim to discretize the continuous latent embeddings $\mathbf{z} \in \mathbb{R}^D$ from the SSL encoders  using RVQ~\cite{zeghidour2021soundstream, defossez2022high}. RVQ approximates $\mathbf{z}$ through a cascade of $L$ quantization layers. At each stage $l \in \{1, \dots, L\}$, the quantizer $Q_l$ identifies the nearest codeword to represent the residual $\mathbf{r}_{l-1}$ from the previous stage, where $\mathbf{r}_0 = \mathbf{z}$ and $\mathbf{r}_l = \mathbf{r}_{l-1} - Q_l(\mathbf{r}_{l-1})$. The reconstructed embedding is given by:
$
\hat{\mathbf{z}} = \sum_{l=1}^{L} Q_l(\mathbf{r}_{l-1})$, with bitrate controlled by number of stages $L$ and codebook size $K$.

We discretize SSL embeddings using RVQ~\cite{zeghidour2021soundstream}.
For RQ1--3 we use $L{=}24$ layers with $K{=}2$ for emotion2vec and $K{=}1024$ for HuBERT/WavLM.
For RQ4 we sweep $(L,K)$ over 7 configurations (Table~\ref{tab:rq4-table}).

\vspace{-3pt}
\subsection{Evaluations \& Metrics}
\vspace{-5pt}

\noindent\textbf{Reconstruction fidelity and primary emotion retention (RQ1 \& RQ2).}
We evaluate how well quantized representations preserve the original continuous embeddings using two metrics: (i) \emph{cosine similarity}, which measures geometric distortion between continuous and quantized representations; and (ii) \emph{primary emotion recall}, which measures the extent to which emotion recognition performance is retained after quantization.

Primary emotion recall is computed with encoder-specific classifiers: the official emotion2vec pretrained linear head (restricted to four emotions) and in-domain, dataset-specific linear probes for HuBERT/WavLM following~\cite{yang21c_interspeech}.
In RQ2, we additionally report normalized codebook entropy for measuring code utilization balance,
\vspace{-10pt}
\begin{equation}
\tilde{H} = \frac{-\sum_{i=1}^{K} p_i \ln p_i}{\ln K},
\end{equation}
\vspace{-0pt}
where $K$ is the codebook size and $p_i$ is the empirical usage of code $i$; $\tilde{H}{=}0$ indicates codebook collapse and $\tilde{H}{=}1$ uniform utilization.

\noindent\textbf{Soft distribution preservation (RQ3).}
We assess how well quantization preserves emotional distributions by comparing the predicted softmax outputs from models using quantized representations against ground-truth annotator vote distributions. We use \textit{Jensen–Shannon (JS) divergence} to measure distributional discrepancy and \textit{Top-2 Set Accuracy}, which assigns a score of 1 if the two most prominent emotions match between prediction and ground truth (order-invariant), and 0 otherwise.

\noindent\textbf{Downstream emotion recognition (RQ4).}
We evaluate downstream SER performance using Macro-F1 of primary emotion.

\vspace{-5pt}

\section{Results and Findings}

\begin{figure}[t]
  \centering
  \includegraphics[width=0.98\linewidth]{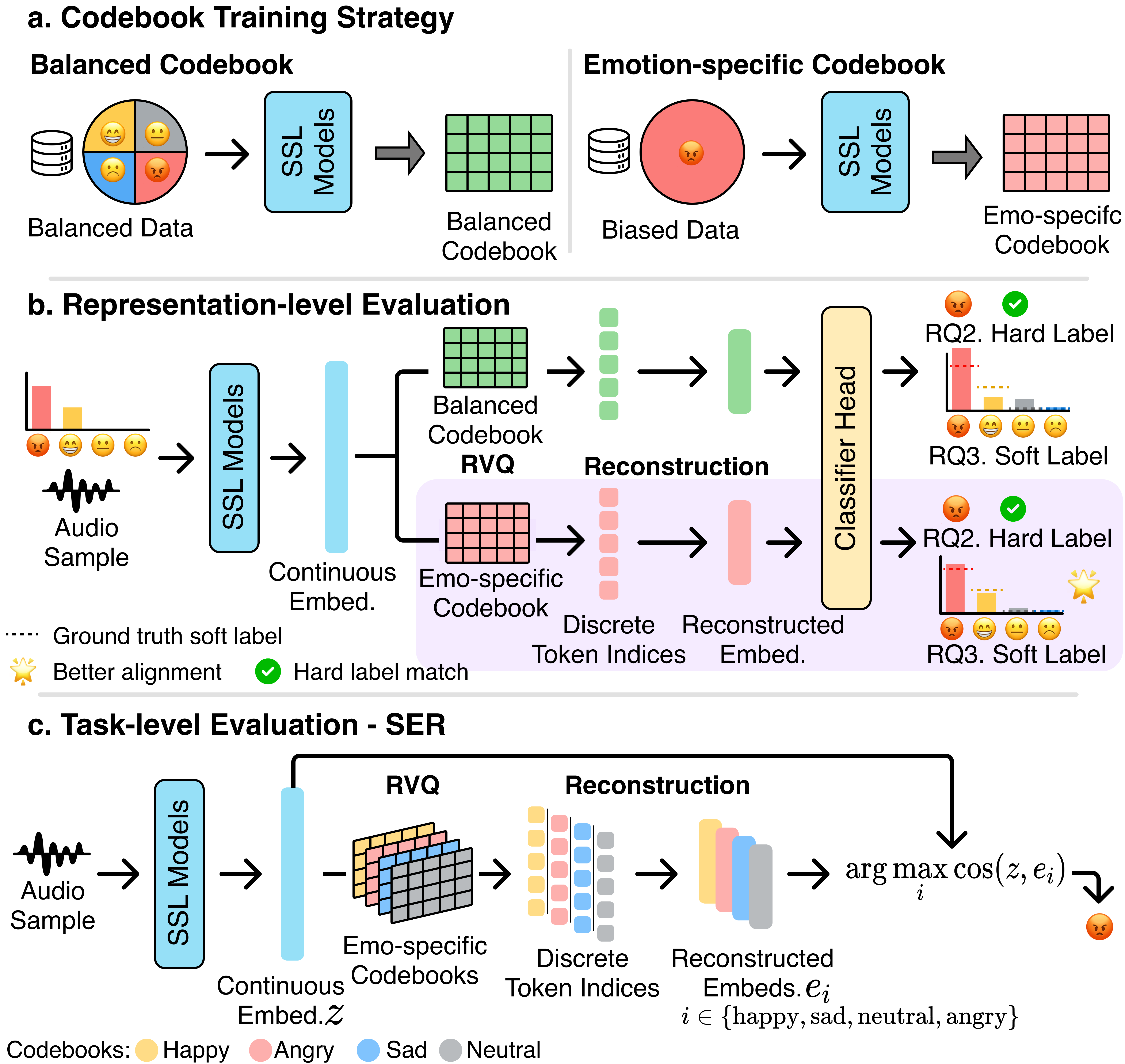}
  \caption{\textbf{Overview of the pipeline}. 
  (a)~Balanced and emotion-specific codebooks.
  (b)~Representation-level evaluation: layer-wise degradation
  analysis (RQ1), primary emotion retention (RQ2), and soft
  distribution fidelity (RQ3).
  (c)~Task-level evaluation: downstream SER via routed
  quantization (RQ4).}
  \vspace{-10pt}
  \label{fig:main}
  \vspace{-5pt}
\end{figure}

\vspace{-2pt}
\subsection{RQ1: How is emotional information structured and degraded under RVQ across different SSL architectures?}
\vspace{-2pt}
We first characterizing how affective information behaves
using standard RVQ compression.
\vspace{-12pt}
\paragraph*{Setup.}

 We evaluate the preservation of emotional information by comparing the original SSL embeddings $z$ against their quantized $\hat{z}$ across all RVQ layers, aiming to reveal how the emotion information degrades. We report the cosine similarity and primary emotion recall for each RVQ layer $L$. 

 As shown in Figure~\ref{fig:main}(a), all codebooks for this analysis are trained on emotion-balanced data with equal numbers of examples per emotion class to isolate the impact of data distribution.

\begin{figure}[t]
    \centering
    
    \begin{subfigure}{\linewidth}
        \centering
        \includegraphics[width=1.0\linewidth]{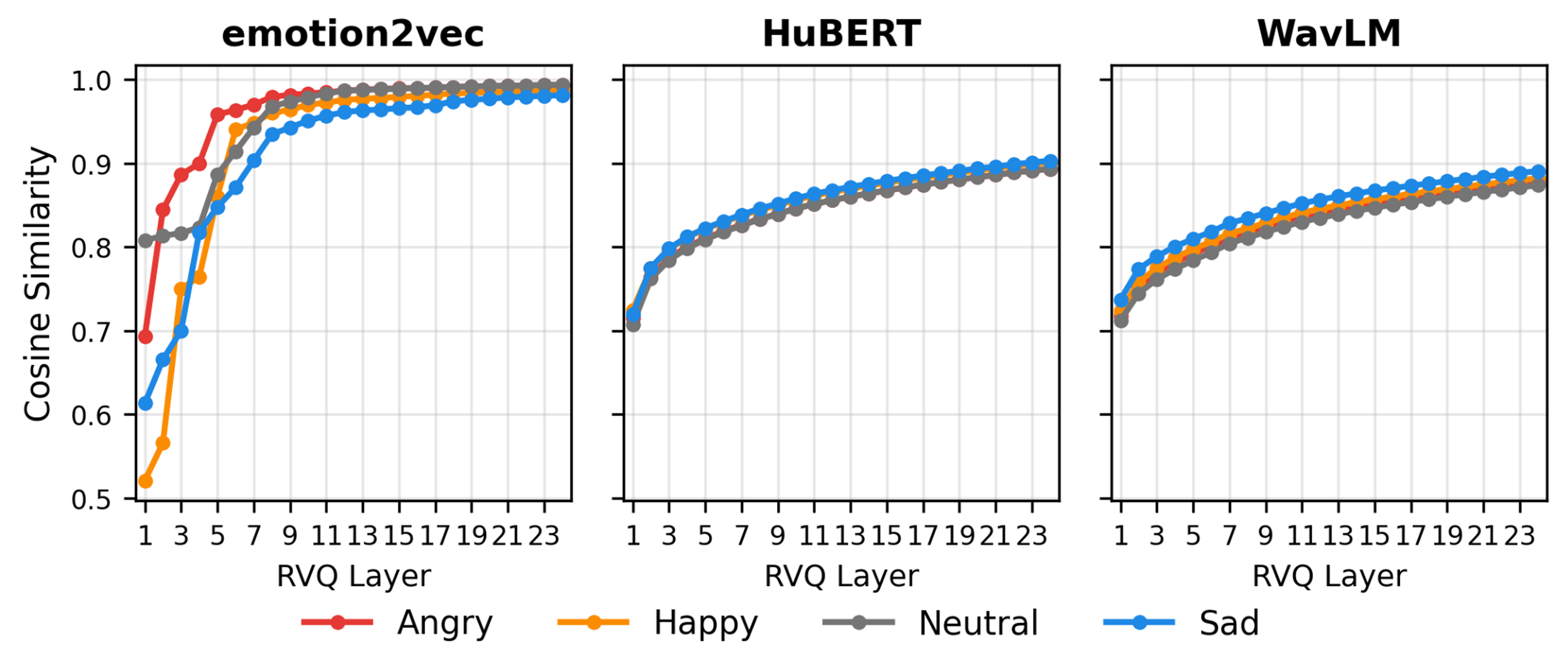}
        \caption{Reconstruction cosine similarity}
        \label{fig:rq1_cossim}
    \end{subfigure}
    
    
    \begin{subfigure}{\linewidth}
        \centering
        \includegraphics[width=1.0\linewidth]{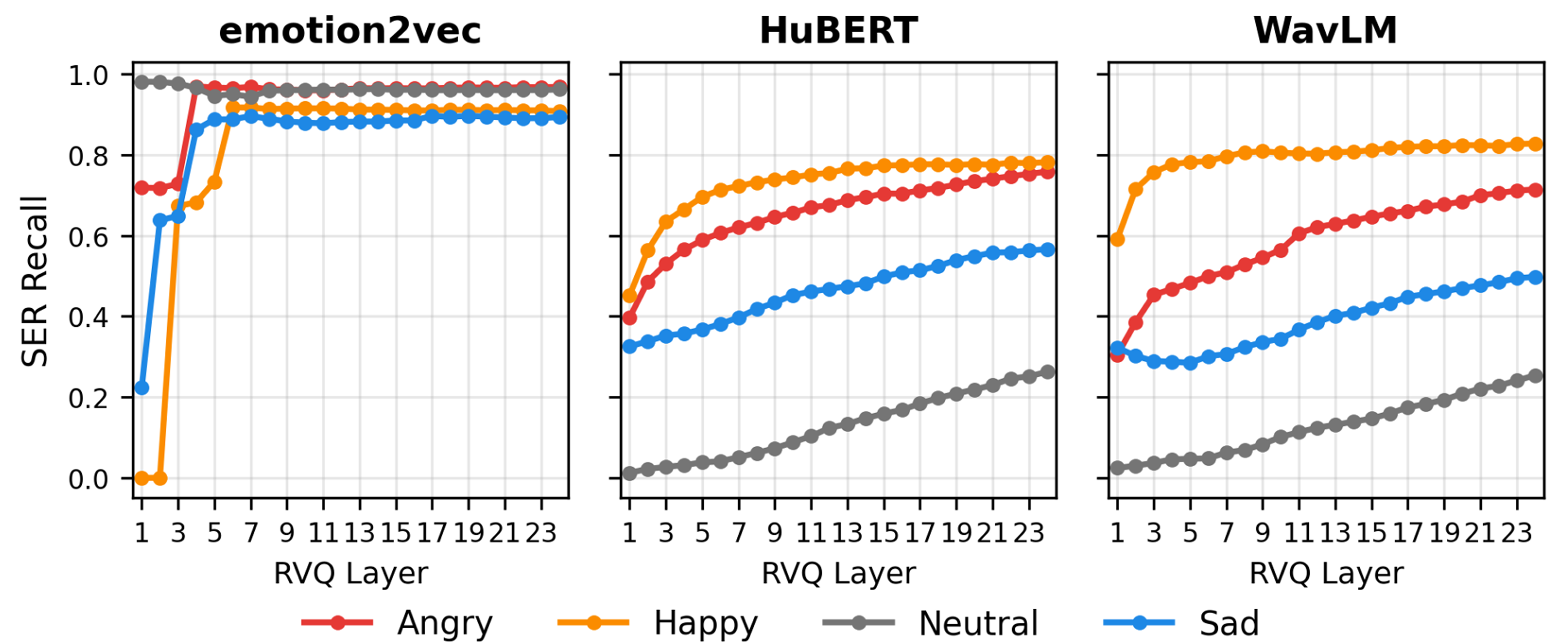}
        \caption{SER primary emotion recall}
        \label{fig:rq1_ser}
    \end{subfigure}
    \vspace{-15pt}
    
    \caption{Reconstruction fidelity (cosine similarity, top) and primary emotion recall (bottom) versus quantization depth under RVQ for three different SSL frontends.}
    \vspace{-10pt}
    \label{fig:rq1_results}
    \vspace{-8pt}
\end{figure}

\vspace{-10pt}

\paragraph*{Findings.}
\textbf{i)~Emotional information is most fragile at low bitrates. }
Across all SSL frontends, Fig.~\ref{fig:rq1_results} shows a consistent trend: both reconstruction cosine similarity and primary emotion recall improve monotonically as the number of RVQ layers increases. The severe performance drop at the first 3–4 layers indicates that strong compression disproportionately removes affective cues. Futhermore, the per-emotion curves diverge under coarse quantization, revealing that some emotions are intrinsically more vulnerable than others. This implies that affective information is not evenly spread in the embedding space, and that early RVQ stages selectively collapse certain emotional subspaces.

\textbf{ii)~Degradation patterns differ between SSLs.}
While all front-ends exhibit monotonic improvement as quantization depth $L$ increases, emotion2vec preserves affective cues more robustly across all layers, exhibiting smaller inter-emotion gaps compared to general-purpose SSLs and maintaining higher primary emotion recall even at low bitrates. In contrast, HuBERT and WavLM show larger early degradation and slower recovery. Notably, neutral becomes the most vulnerable emotion in these general-purpose SSLs, while it remains relatively stable in emotion2vec.
These results suggest that robustness to discretization is shaped by the SSL pretraining objective. Emotion-targeted models likely encode affective information within more compact subspaces, making it more resilient to coarse quantization. General-purpose SSLs are pretrained using masked acoustic prediction, where emotional cues are not explicitly targeted and are instead distributed diffusely across the representation space. The weaker preservation of neutral speech in these models may reflect the lack of strongly discriminative acoustic patterns, rendering its representation less well-defined and more vulnerable to quantization noise.

\vspace{-5pt}

\subsection{RQ2: Can emotion-specific codebook training improve affective preservation?}
\vspace{-2pt}
\label{section-rq2}

Following RQ1's evidence of uneven affective decay, we investigate whether the quantizer can be made more emotion-aware. As shown in Figure~\ref{fig:main}(b), we hypothesize that training codebooks on single-emotion data will shift quantization centroids toward emotion-relevant regions of the representation space, enabling more emotion-aligned tokenization and reducing affective information loss under compression. 
\vspace{-5pt}

\label{sec:rq2}

\begin{figure*}[t]
\centering
\begin{subfigure}[t]{0.37\textwidth}
  \centering
  \includegraphics[width=\linewidth]{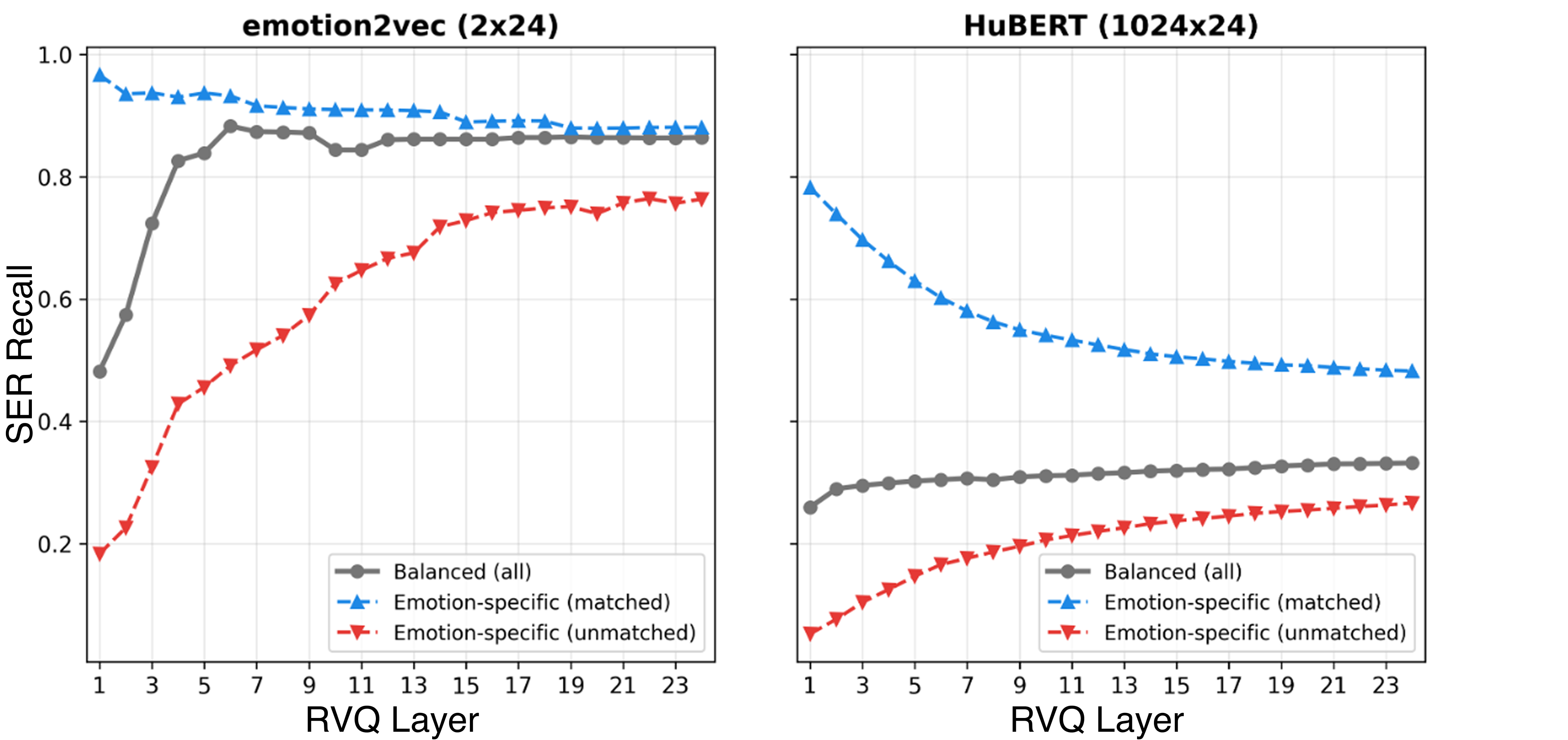}
  \caption{SER primary recall}
  \label{fig:rq2-a}
\end{subfigure}\hfill
\begin{subfigure}[t]{0.63\textwidth}
  \centering
  \includegraphics[width=\linewidth]{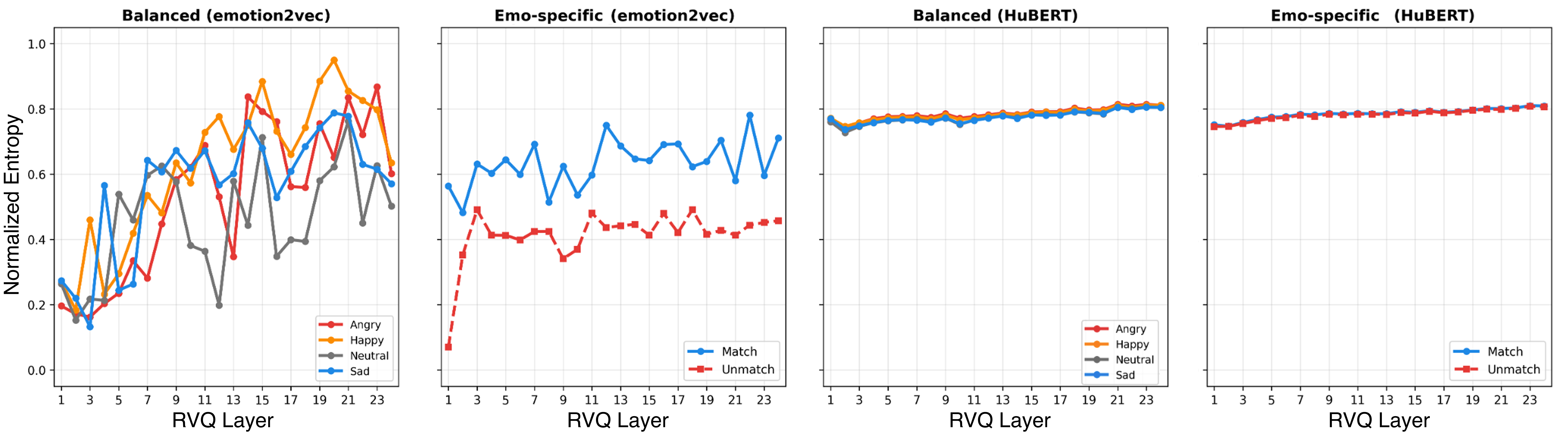}
  \caption{Normalized Shannon entropy}
  \label{fig:rq2-b}
\end{subfigure}
\vspace{-5pt}
\caption{RQ2: affective retention (left) and codebook utilization (right) for balanced and emotion-specific quantization.}
\label{fig:rq2-results}
\vspace{-5pt}
\end{figure*}

\vspace{-5pt}

\paragraph*{Setup.}

We compare a \textit{balanced codebook} trained on emotion-uniform data with \textit{emotion-specific codebooks} trained on single-emotion data (Fig.~\ref{fig:main}(b)), while keeping the total number of training utterances constant.
During evaluation, we test emotion-specific codebooks in two ways: a \textit{matched} setting, where the input emotion is the same as the emotion the codebook was trained on, and an \textit{unmatched} setting, where they differ. The matched case shows how well a codebook can preserve its target emotion, while the unmatched case checks how much that codebook blocks or distorts other emotions.

\vspace{-13pt}

\paragraph*{Findings.}
\textbf{i)~Emotion-specific codebooks improve target emotion preservation under strong compression.}
As shown in Fig.~\ref{fig:rq2-a}, 
emotion-specific codebooks consistently outperform the balanced baseline for \textit{matched} emotions in primary emotion recall, with the largest gains observed at low bitrates. This suggests that training codebooks on single-emotion data helps quantization tokens better capture emotion-relevant acoustic patterns. The improvement is most pronounced for the emotion-targeted frontend (\textit{emotion2vec}), whose matched recall remains stable across RVQ depth. In contrast, general-purpose SSL (represented here by HuBERT, which exhibits trends nearly identical to WavLM) models exhibit gradual degradation at higher bitrates, consistent with deeper RVQ layers encoding fine-grained phonetic variability that is weakly related to emotion. In the \textit{unmatched} condition, early RVQ layers produce substantially lower recall than in the matched setting. As bitrate increases, unmatched performance recovers toward the balanced baseline, confirming that emotion-specific codebooks impose a selective bottleneck under strong compression while retaining sufficient flexibility at higher bitrates.

\textbf{ii)~Gains stem from capacity reallocation rather than code collapse.}
To investigate the mechanism of this improvement, in Fig.~\ref{fig:rq2-b}, we analyze the normalized Shannon entropy ($\tilde{H}$) of codebook utilization across each RVQ layer. Under matched conditions, emotion-specific codebooks keep entropy high across layers, meaning they actively use many different codes. This suggests that the gains in primary emotion recall come from reallocating capacity toward emotion-relevant patterns, rather than simply collapsing onto a few frequent codes.

\vspace{-4pt}

\subsection{RQ3: Does emotion-specific quantization preserve fine-grained affective structure under compression?}
\vspace{-5pt}

\label{sec:rq3}
Building on the categorical improvements in RQ2, we investigate whether emotion-specific codebooks preserve fine-grained affective structures beyond discrete labels. Since real-world affect is inherently mixed and ambiguous, 

\vspace{-8pt}

\begin{figure*}[t]
\centering
\begin{subfigure}[t]{0.5\textwidth}
  \centering
  \includegraphics[width=\linewidth]{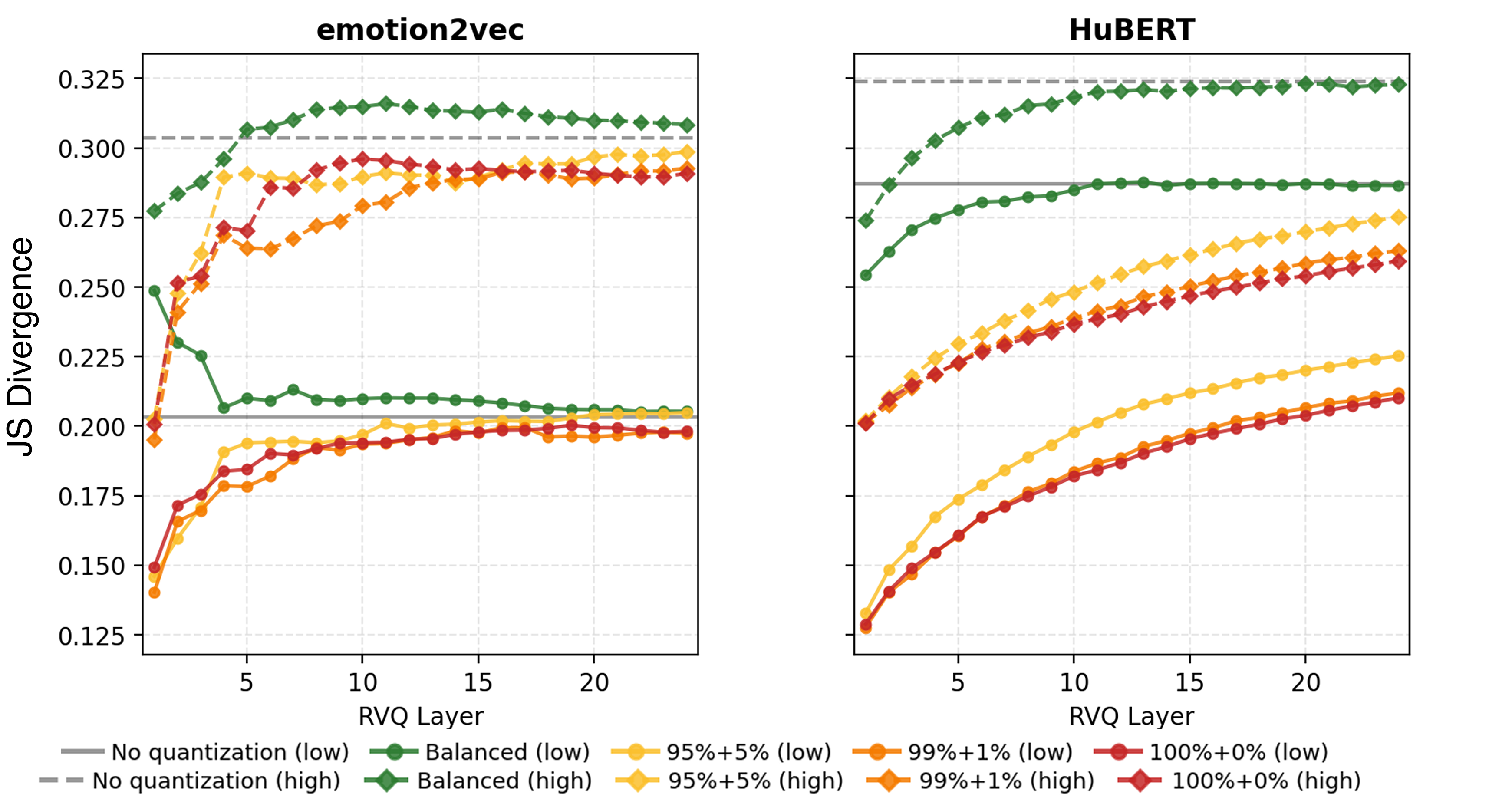}
  \caption{JS divergence}
  \label{fig:rq3-a}
\end{subfigure}\hfill
\begin{subfigure}[t]{0.49\textwidth}
  \centering
  \includegraphics[width=\linewidth]{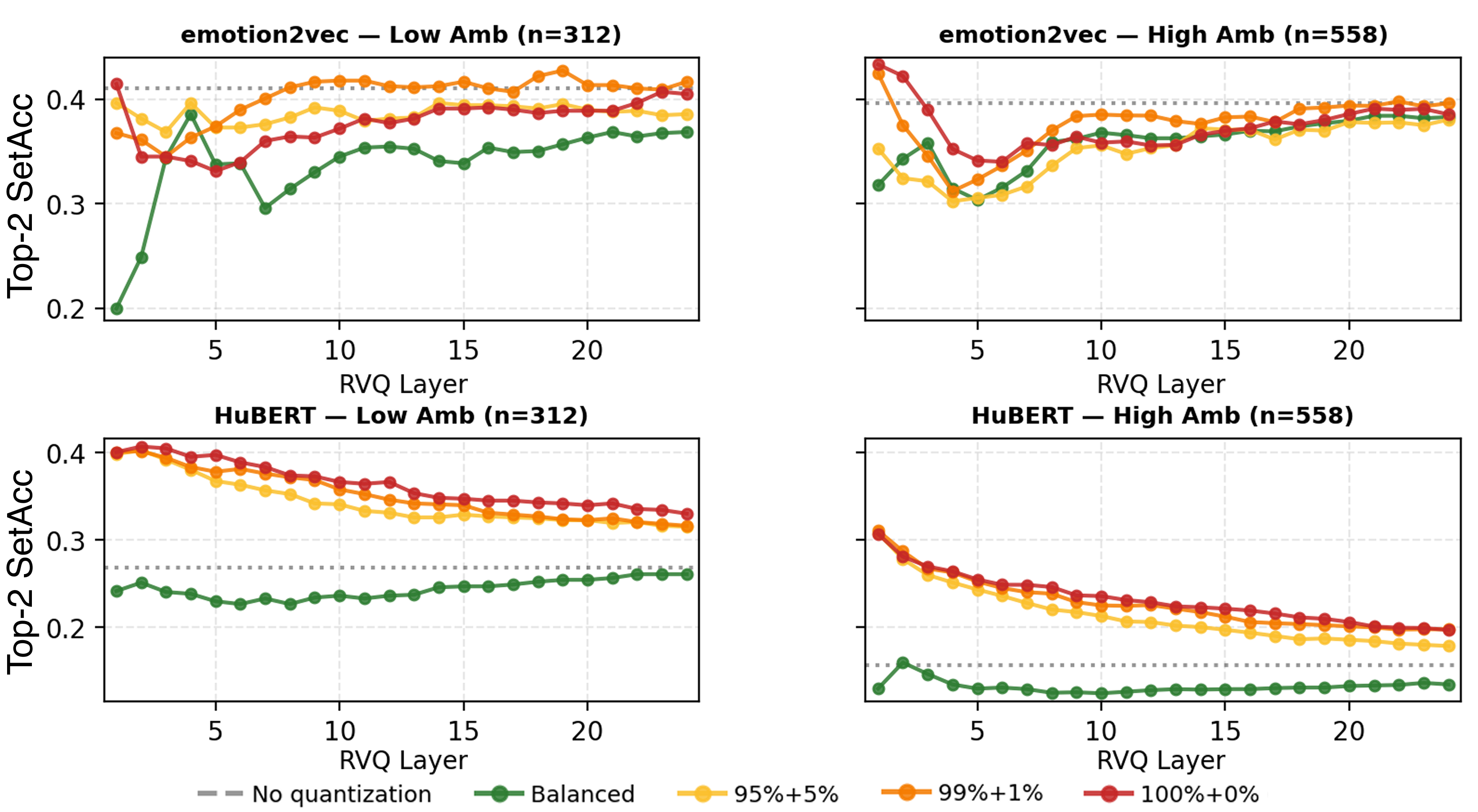}
  \caption{Top-2 set accuracy}
  \label{fig:rq3-b}
\end{subfigure}
\vspace{-10pt}
\caption{RQ3 evaluation: emotion distribution matching (left) and top2 emotion recall (right) for varying codebook training strategies.}
\vspace{-10pt}
\label{fig:rq3-results}
\vspace{-5pt}
\end{figure*}

\vspace{-10pt}

\paragraph*{Setup.}
We evaluate how well emotion-specific codebooks preserve full emotion distributions by comparing the predicted soft labels to the ground-truth soft labels (e.g., 30\% happy, 70\% surprised). For each utterance, we feed the quantized representations into a classifier to obtain a predicted probability distribution over emotions, then compare this distribution with the annotator-derived distribution using Jensen–Shannon (JS) divergence and Top-2 set accuracy using the IEMOCAP corpus (as shown in Fig.~\ref{fig:main}(b)). 

To evaluate robustness under varying affective complexity, we stratify the data into two subsets:
\vspace{-2pt}

\begin{itemize}[leftmargin=1.2em,noitemsep]
\item \textit{Low-ambiguity}: utterances where the primary emotion is dominant (vote share $>50\%$).
\item \textit{High-ambiguity}: utterances with mixed affect where no emotion receives a majority (vote share $\leq50\%$).
\end{itemize}

To examine how codebook training composition affects distribution preservation, we evaluate three schemes: (i) a \emph{balanced} codebook, (ii) \emph{emotion-specific} codebooks $100+0$, and (iii) \emph{emotion-biased with mixed ratios} $A+(100-A)$, where $A\%$ of samples belong to the target emotion and the remainder are drawn uniformly from other emotions.
The total number of training utterances is held constant across all conditions.

\vspace{-15pt}

\paragraph*{Findings.}
\textbf{i) Emotion-specific training improves the preservation of soft affective distribution.}
As shown in Fig.~\ref{fig:rq3-a}, across both emotion2vec and HuBERT (with WavLM exhibiting a similar trend), emotion-specific and emotion-biased codebooks consistently achieve lower JS divergence than the balanced baseline in both low- and high-ambiguity settings. This improved distributional fidelity is further supported by the Top-2 set accuracy results in Fig.~\ref{fig:rq3-b}. This indicates that co-orruring emotions are better preserved under specialized quantization. Collectively, these results suggest that emotion-specific and emotion-biased training reallocates quantization capacity toward within-emotion variation, enabling preservation of the full probabilistic affective structure rather than only the dominant category.

\textbf{ii) Mild impurity regularizes emotion-aware quantization.}
To identify the optimal configuration for preserving fine-grained structure, we analyze the impact of training ratios ($A\% + (100-A)\%$). In Fig.~\ref{fig:rq3-b}, while purely specialized training (100+0) performs well, lightly mixed regimes (95+5 and 99+1) are consistently competitive and sometimes superior, particularly for emotion2vec. 

This suggests that a small degree of cross-emotion exposure during codebook training improves coverage of emotion boundaries, reducing
over-specialization and stabilizing preservation of subtle
affective cues.

\vspace{-5pt}

\vspace{-3pt}
\subsection{Downstream Utility: Emotion-aware SER with Routed Quantization}
\vspace{-5pt}
\label{sec:downstream_ser}

Leveraging the selective bottleneck identified in Sec.~\ref{section-rq2}), we hypothesize that reconstruction fidelity under specialized codebooks serves as an inherent discriminative signal. We thus evaluate the practical utility of these representations via a lightweight primary emotion recognition pipeline using similarity-based routing.

\vspace{-15pt}

\paragraph*{Setup.} 
As illustrated in Fig.~\ref{fig:main}(c), given an input SSL embedding $z$ of emotion2vec, we calculate its reconstruction embeddings $\mathbf{e}_i$ through each emotion-specific codebooks. The core routing mechanism, \textbf{Emo-Q}, emotion-aware quantization assigns the label of the codebook with the highest cosine similarity: $\hat{y} = \arg\max_i \cos(z, e_i)$, where $\mathbf{e}_i$ represents reconstructed vector from the $i$-th emotion-specific codebook.

\noindent We compare \textbf{Emo-Q} against two reference conditions:
\vspace{-3pt}
\begin{itemize}
\item \textbf{Baseline}: Continuous emotion2vec embeddings processed by the official downstream SER classifier.
\item \textbf{Bal} (Balanced): Discrete representations derived from a balanced codebook at the optimal RVQ layer, subsequently processed by the same SER classifier.
\end{itemize}
\vspace{-3pt}
\noindent Both 100+0 and 99+1 codebook training ratios (Section~\ref{sec:rq3}) are evaluated. Results are reported as Macro~$F_1$ change ($\Delta$) relative to the continuous baseline, averaged across four OOD test sets.

\vspace{-10pt}

\paragraph*{Findings.}
Table~\ref{tab:rq4-table} shows that \textbf{Emo-Q} similarity-based routing consistently improves SER performance under quantization and often surpasses the continuous baseline. The largest gains are observed for the 8$\times$32 configuration. Across most settings, the 99+1 mixed training ratio yields more stable improvements than the pure 100+0 setting, suggesting that slight impurity regularizes the emotion-specific codebooks and mitigates over-specialization. From an efficiency perspective, Emo-Q (8$\times$32) enables over a 500$\times$ reduction in transmission bitrate compared to transmitting continuous SSL embeddings ($\approx$1.4(avg.) kbps vs. $\approx$1.2 Mbps), while maintaining or even improving classification accuracy through a zero-parameter routing mechanism.

The primary practical trade-off lies in storage. Emo-Q requires maintaining four emotion-specific codebooks ($\approx$6.4 MB) instead of a single balanced codebook used in standard quantization. However, this overhead remains negligible relative to the 350 MB SSL backbone, while enabling both efficient discretization and improved paralinguistic preservation.

\begin{table}[t]
\centering
\small
\caption{Primary emotion Macro-$F_{1}$ improvement ($\Delta$) over the continuous baseline for different codebook configurations (Codebooks $\times$ Entries).}
\vspace{-5pt}
\setlength{\tabcolsep}{2pt} 
\renewcommand{\arraystretch}{1.25}

\resizebox{\columnwidth}{!}{%
\begin{tabular}{lcccccccc}
\toprule
\textbf{Config.} & $8{\times}32$ & $8{\times}64$ & $8{\times}128$ & $32{\times}2$ & $32{\times}4$ & $64{\times}2$ & $128{\times}2$ \\
\midrule
\textbf{Bal} & $-0.09$ & $+0.02$ & $+0.02$ & $-0.98$ & $-0.53$ & $-1.00$ & $+0.17$ \\
\textbf{Emo-Q\textsubscript{(100)}} & $+0.95$ & $+0.97$ & $+0.66$ & $+0.51$ & $\mathbf{+1.18}$ & $\mathbf{+0.67}$ & $\mathbf{+0.77}$ \\
\textbf{Emo-Q\textsubscript{(99)}}  & $\mathbf{+1.57}$ & $\mathbf{+1.45}$ & $\mathbf{+1.20}$ & $\mathbf{+0.57}$ & $+0.25$ & $-0.18$ & $-0.18$ \\
\bottomrule
\end{tabular}
} 
\label{tab:rq4-table}
\vspace{-20pt}
\end{table}

\vspace{-3pt}

\section{Conclusion}
This work examined how residual vector quantization reshapes emotional information in self-supervised speech representations, and how quantizers can be redesigned to better preserve affect. While aggressive RVQ compression disproportionately harms emotion, we introduced emotion-aware codebooks that improve both categorical and distributional preservation and leverage them for improved SER performance. Our findings argue that emotion-aware discretization should be treated as a core design choice in speech codecs and tokenizers.

\section{Generative AI Use Disclosure}
Generative AI tools were employed solely for improving the linguistic quality and clarity of this manuscript. While the AI provided suggestions for phrasing and sentence structure, the conceptual framework and all technical findings were developed by the authors, who take full responsibility for the integrity and accuracy of the work.
\bibliographystyle{IEEEtran}
\bibliography{mybib}

\end{document}